\documentclass{jkas}
\usepackage{comment}
\usepackage{amsmath}

\def\year{2020} 
\def\month{November} 
\def\volume{00} 
\def\issue{0} 
\def\beginpage{1} 
\def\endpage{99} 
\def\received{September 30, 2020} 
\def\accepted{, 2020} 
\date{Received \received; accepted \accepted}

\title{
3D Grid-Based Monte Carlo Code for Radiative Transfer through Raman and Rayleigh Scattering with Atomic Hydrogen
        -- STaRS
}

\author[1,2]{Seok-Jun Chang}
\author[1]{Hee-Won Lee}

\affil[1]{Department of Physics and Astronomy, Sejong University, Seoul, Korea; \email{csj607@gmail.com}}
\affil[2]{Korea Astronomy and Space Science Institute,
Daejeon, Korea}

\def\corrauthor{
Seok-Jun Chang
}

\def\runningauthor{
Chang \& Lee
}

\def\runningtitle{
Monte Carlo Radiative Transfer for Raman Scattering - STaRS 
}

\def\keywords{
Radiative Transfer --- Scattering --- Polarization
}

\def\abstracttext{
Emission features formed through Raman scattering with atomic hydrogen provide unique and crucial information to
probe the distribution and kinematics of a thick neutral region illuminated by a strong far UV emission source.
We introduce a new 3 dimensional Monte-Carlo code in order to describe the radiative transfer of line
photons that are subject to Raman and Rayleigh scattering with atomic hydrogen.
In this code entitled "{\bf S}ejong Radiative {\bf T}r{\bf a}nsfer through {\bf R}aman and
Rayleigh {\bf S}cattering ({\it STaRS}), each photon is traced until escape with a tag attached carrying information including the position, direction, wavelength, and polarization.
The thick neutral scattering region is divided into numerous cells with each cell being characterized by its velocity and
density, which ensures huge flexibility of the code in analyzing Raman-scattered features formed in a neutral region
with complicated kinematics and density distribution.
As a test of the code, we revisit the formation of Balmer wings through Raman scattering of
far UV continuum near Ly$\beta$ and Ly$\gamma$ in a static neutral region. An additional check is made
to investigate Raman scattering of O VI in an expanding neutral medium. We find fairly good
agreement of our results with previous works, demonstrating the capability of dealing with radiative transfer modeling that can be applied to spectropolarimetric imaging observations of various objects including symbiotic stars, young planetary nebulae, and active galactic nuclei.
}

\begin{document}
\jkashead 


\section{Introduction} \label{sec:intro}

Scattered radiation conveys special information regarding both the emission and scattering regions.
In particular, linear polarization may develop in an anisotropic scattering geometry and the relative motion
between the emission and scattering regions gives rise to complicated line profiles including multiple-peak 
structures and broad wings. An excellent example is provided by spectropolarimetric observations
of Seyfert 2 galaxies exhibiting broad lines in the linearly polarized fluxes, lending strong support
to the unification model of active galactic nuclei
\citep{miller90, tran10}.

Interaction of electromagnetic radiation with an atomic electron may be classified into Rayleigh
and Raman scattering \citep[e.g.][]{sakurai67}. Rayleigh scattering refers to the elastic process where the scattered photon
has the same wavelength as the incident one. Otherwise, we have Raman scattering in which the initial and final electronic states differ so that the scattered photon emerges with the energy difference, which is enforced by the law of
energy conservation. Raman spectroscopy is particularly useful in revealing the complicated energy level structures
of a molecule in chemistry.

In astrophysics, 
\cite{nussbaumer89} provided a pioneering discussion on Raman scattering with atomic hydrogen, introducing
a new spectroscopic tool to diagnose gaseous emission nebulae including symbiotic stars and active galactic
nuclei. Relevant Raman scattering processes start with a far UV photon more energetic than Ly$\alpha$
incident on an hydrogen atom in the ground state and end with an outgoing photon with energy less than 
that of Ly$\alpha$ leaving behind the hydrogen atom in the $2s$ state.
They introduced basic atomic physics of Rayleigh and Raman scattering illustrating the cross sections
and presented a number of candidate far UV spectral lines that may result in detectable Raman-scattered features
\citep[e.g.,][]{saslow69}. 
One notable point is that the branching ratio of Raman and Rayleigh scattering of far UV radiation near Ly$\beta$
is approximately 0.15 so that most Ly$\beta$ photons are Rayleigh (or resonantly) scattered several times before
they are converted into H$\alpha$ photons to escape from the thick neutral region.

\cite{schmid89} identified the broad emission features at 6825 \AA~and 7082 \AA~ found 
in about half of symbiotic stars as Raman scattered features of  O VI $1032$ and $1038$ \AA~ emission 
lines, respectively \citep{akras19}. Symbiotic stars are wide binary systems composed of a hot white dwarf and 
a mass-losing red giant. Hydrodynamical studies suggest that 
some fraction of slow stellar wind is gravitationally captured to form an accretion disk \citep{devalborro17,chen17,saladino18}, where
O VI$\lambda\lambda$1032 and 1038 are important coolants.

Considering very small Raman scattering cross sections $\sim 10^{-23}
{\rm\ cm^2}$ for O VI$\lambda\lambda$1032 and 1038 doublet lines, the operation of Raman scattering 
requires a special condition that a very thick neutral
region is present in the vicinity of a strong O VI emission region. This special condition is ideally met in symbiotic
stars, where a thick neutral region surrounding the giant component is illuminated by strong far UV radiation
originating from the nebular region that may be identified with the accretion flow onto the hot component.

Raman-scattered O VI features in symbiotic stars exhibit
complicated profiles with multiple peaks separated by $\sim 30-50{\rm\ km\ s^{-1}}$ indicative
of the O VI emission regions with physical dimension of $\sim 1{\rm\ au}$ \citep[e.g.][]{shore10,heo15, lee19}.
Because Raman and Rayleigh scattering sufficiently off resonance shares the same scattering phase function
as Thomson scattering\citep{schmid95,yoo02,chang17}, strong linear polarization may develop in an anisotropic scattering geometry.
\cite{harries96} conducted spectropolarimetric observations of many symbiotic stars to show that Raman-scattered
O VI features are strongly polarized. They also found that most Raman-scattered O VI features show polarization
flip in the red wing part, where the polarization develops nearly perpendicularly to the direction along which
the main part is polarized. \cite{lee99} proposed that the polarization flip is closely associated with the bipolar
structure of symbiotic stars \citep[see also][]{heo16}.

Raman scattering plays an interesting role of redistributing far UV radiation near Ly$\beta$ and Ly$\gamma$
into near H$\alpha$ and H$\beta$, respectively. He~II being a single electron ion with $Z=2$, the transition 
to $n=2$ from an energy level from $n=2k, k>1$ gives rise to emission lines with wavelengths slightly shorter than
those of H I Lyman series $k \to 1$, for which the cross sections for Rayleigh and Raman scattering
are conspicuously large. \cite{vangroningen93} found Raman-scattered He~II features near H$\beta$ in 
the symbiotic nova RR~Telescopii. \cite{pequignot97} discovered the same spectral feature
in the young planetary nebula NGC~7027, which constitutes the first discovery of a spectral feature formed through Raman
scattering with atomic hydrogen in planetary nebulae.
Subsequently, Raman-scattered He~II at $6545$ \AA~ has been detected in several symbiotic stars \citep{birriel04,jung04,sekeras15}
and in the young planetary nebulae NGC~6302, IC~5117, NGC~6790, NGC~6886, and NGC~6881 \citep{groves02, lee01,lee06,lee09,letter20}.

It is particularly notable that the
case B recombination theory allows one to deduce the strengths of incident far UV He~II lines \citep{storey95}, yielding faithful
estimates of Raman conversion efficiencies. This is extremely useful in the measurement of
H~I content in symbiotic stars and young planetary nebulae, which makes Raman spectroscopy a totally new
approach to probing the mass loss processes occurring in the late stage of stellar evolution
\citep{lee06,choi20}.
Furthermore, H$\alpha$ and H$\beta$ in symbiotic stars and young planetary nebulae often display fairly 
extended wings that may indicate the
presence of fast tenuous stellar wind \citep[e.g.,][]{arrieta02}. Broad wings around Balmer lines may also arise via Raman scattering far-UV 
continuum around Lyman lines, which requires further investigation \citep{lee00a,yoo02,chang18}.

Additional examples include Raman scattering of C II$\lambda\lambda$1036 and 1037 forming 
optical features at 7023 \AA\ and 7054 \AA, which
were reported in the symbiotic nova V1016~Cygni by \cite{schild96}. \cite{dopita16} investigated
the H II regions in the Orion Nebula (M42) and five H II regions in the Large and Small
Magellanic Clouds to discover Raman-scattered features at 6565 \AA\ and 6480 \AA\ formed through Raman scattering of O I $\lambda$1025.76 and Si II $\lambda$1023.70, respectively.

The Monte Carlo approach is an efficient numerical technique to describe radiative transfer in various regions having
	 a dust component\citep[e.g.][]{celnikier74,seon15} and a molecular component\citep[e.g.][]{brinch10}.
	 A similar approach has been applied to radiative transfer of an  
	electron scattering \citep[e.g.][]{angel69,seon94}. Ly$\alpha$ deserves special attention being
	characterized by large scattering optical depth \citep[e.g.][]{eide18,seon20}.

\cite{schmid92} performed  Monte-Carlo simulations to investigate the formation of Raman-scattered O VI
features in an expanding H~I region with an assumption that a given line photon has an invariant scattering cross 
section as it propagates through the H~I medium. Similar studies were presented by \cite{lee97b}, who
adopted a density matrix formalism to determine the physical information of scattered radiation including polarization.
\cite{chang15} investigated the formation of broad Balmer wings near H$\alpha$ and H$\beta$
in the unification scheme of active galactic nuclei and presented quantitatively
the asymmetry of the wings formed in neutral regions with extremely high H I column density $\sim 10^{23} \rm\ cm^{-2}$.

In this paper, we introduce a new grid-based Monte Carlo code entitled "Sejong Radiative Transfer for Rayleigh 
and Raman Scattering ({\it STaRS}), 
We also present our test of the code by revisiting the formation of Balmer wings and Raman O VI features
through Raman scattering with atomic hydrogen.

\section{Grid Based Monte Carlo Simulation}\label{sec:method}

\begin{figure}
	\includegraphics[width=90mm]{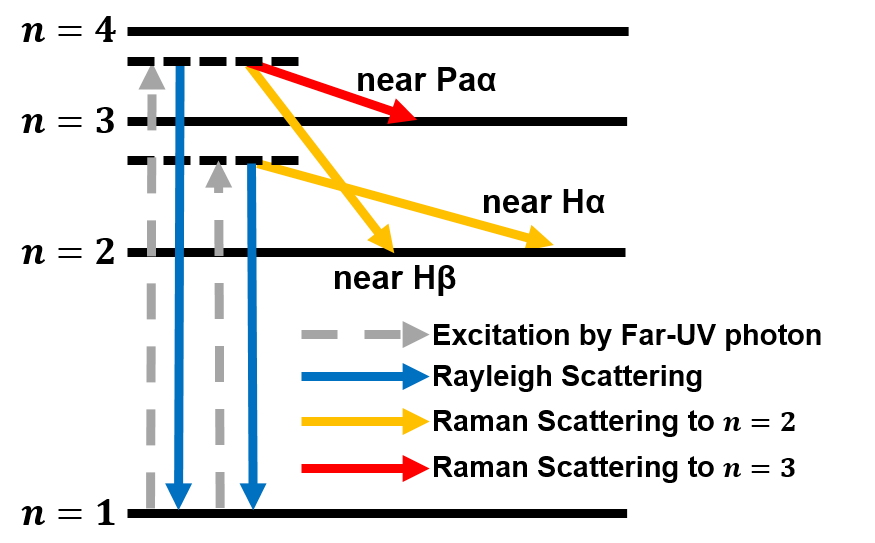}
	\caption{Schematic illustrations of energy levels and electronic transitions associated with Raman and Rayleigh scattering with a hydrogen atom of far UV electromagnetic radiation.
	}
	\label{fig:scattering}
\end{figure}

In this section, we describe 'STaRS' and discuss the basic atomic physics of Rayleigh and Raman scattering with atomic hydrogen.
Fig.~\ref{fig:scattering} is a schematic illustration of a few 
representative transitions pertaining to the two types of scattering. Thus far, detected Raman-scattered
features are limited to those associated with the final de-excitation to $2s$ state.
The second order time-dependent perturbation theory is used to compute the scattering cross sections
known as the Kramers-Heisenberg formula \citep[e.g.,][]{bethe67, sakurai67, saslow69}. 

In our grid based Monte Carlo code, we divide
the region of interest into a large number of small cubes or cells in the Cartesian coordinate system, where each cell is characterized by uniform physical 
properties in the three dimensional space.
Here, the uniform physical properties include
H I number density $n_{{\rm HI},G}$ and the velocity ${\bf v}_{G}$.
We  assign the emissivity $j_{e}$ to each cell and
generate an initial photon using $j_{e}(\lambda_i,x,y,z)$, where $\lambda_i$ is the wavelength of the initial photon.
STaRS is mainly written in {\it Fortran} with Message Passing Interface implemented for parallel computing. 
We also adopt the shared memory technique by {\it intel MPI}.

\begin{figure}
	\includegraphics[width=90mm]{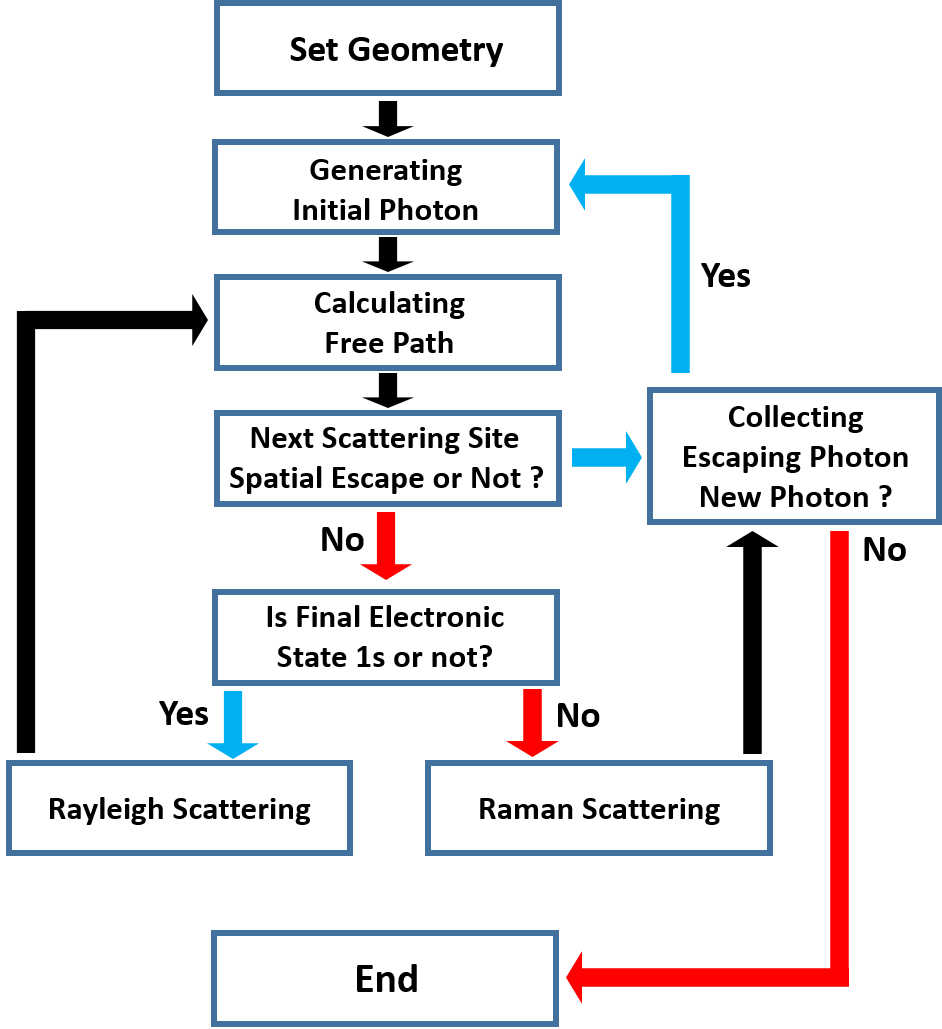}
	\caption{A flow chart of STaRS.
	}
	\label{fig:flow_chart}
\end{figure}

Fig.~\ref{fig:flow_chart} shows a flow chart for radiative transfer simulations using STaRS.
A simulation starts with the setup of the scattering geometry by assigning to each cell appropriate
physical conditions. Initial far UV photons are generated in accordance with our prescription of $j_{e}$.
Each photon is tagged with the information of
the unit wavevector $\bf{\hat k}$, the position vector $\bf r$, the wavelength $\lambda$,
and the density matrix $\rho$ composed of the Stokes parameters, $I$, $Q$, $U$, and $V$. 
The free path $d$ for next scattering position $\bf r'$ is computed by transforming the scattering optical
depth into the physical depth. 

Decision is made whether the photon escapes from the region or is scattered into another direction.
If the next scattering position $\bf r'$ is outside the scattering region, the photon is assumed to
reach the observer as a far UV photon. Otherwise, we generate a new photon using the scattering
phase function for an electric dipole process and determine the scattering type.
We assume that the region is transparent to Raman-scattered photons. The scattered photon escapes 
from the scattering region if the scattering is Raman. Otherwise, the Rayleigh scattered photon
is regarded as an incident photon propagating to a new scattering position.
The procedure is repeated until escape.

In Sec.~\ref{sec:geometry} and \ref{sec:source}, we provide
more detailed descriptions on the scattering geometry and generation of initial far UV photon prescribed
by $j_{e}$. In Sec.~\ref{sec:path}, we describe the computation of a free path $d$.
We discuss the basic properties of the scattered photons in Sec.~\ref{sec:scattering}.

\subsection{Geometry : Scattering Region}\label{sec:geometry}

The medium for radiative transfer through Raman and Rayleigh scattering
corresponds to a thick H I region that is easily found in the slow stellar wind from a red giant \citep{lee97b,lee19}.
No consideration on the thermal motion of neutral hydrogen is given
because the variation of the cross section and the branching ratio in the scale of thermal speed
is negligible \citep{chang15,chang18}.

In our simulation, we divide the scattering region into a large number of cells having a fixed size.
Each cell is identified by a set of three indices $(iX, iY, iZ)$ with $iX, iY$ and $iZ$ running from 1 to
$N_X, N_Y$ and $N_Z$. 
If we denote by $X_{min}$ and $X_{max}$ the range of $x$ coordinates for the scattering region,
the boundary of the $iX$th cell on the $x$ axis is given by
\begin{equation}
X_G(iX) = X_{min} + \frac{X_{max} - X_{min}}{N_X} (iX - 1).
\end{equation}
We also define $Y_G(iY)$ and $Z_G(iZ)$ in a similar way.

Therefore, for any point $P(x,y,z)$ in the cell identified with $(iX,iY,iZ)$, we have 
\begin{eqnarray}
X_G(iX) < x < X_G(iX+1) \\
\nonumber
Y_G(iY) < y < Y_G(iY+1) \\
\nonumber
Z_G(iZ) < z < Z_G(iZ+1).
\end{eqnarray}
Each cell is also identified by its center point, whose coordinates
are given by the following relations
\begin{eqnarray}
X_C(iX) = \frac{X_G(iX) + X_G(iX+1)}{2} \\
\nonumber
Y_C(iY) =  \frac{Y_G(iY) + Y_G(iY+1)}{2} \\
\nonumber
Z_C(iZ) =  \frac{Z_G(iZ) + Z_G(iZ+1)}{2}.
\end{eqnarray}

\subsection{Emission Source : Initial Photon}\label{sec:source}

The operation of Raman scattering with atomic hydrogen requires the coexistence of a strong far UV source 
and a thick neutral region. In the case of symbiotic stars and young planetary nebulae, a strong far UV 
emission region is formed near an accreting white dwarf or a hot central star and a thick neutral region
is also present in association with the mass loss of a giant star. In the simulation,
the emissivity $j_{e}(\lambda,x,y,z)$ is prescribed as a function of wavelength and position.
Regarding $j_{e}$ as the normalized probability density function,
we pick a wavelength $\lambda_i$ and a starting 
position $(x_i,y_i,z_i)$ of an initial photon in accordance with $j_{e}(\lambda,x,y,z)$.

We find the spatial index $(iX,iY,iZ)$ from $(x_i,y_i,z_i)$ to determine the cell that contains the starting
position. For simplicity, we assume that the initial photon is completely unpolarized and that the unit 
wavevector of the initial photon is selected from an isotropic distribution. 
Thus, the initial unit wavevector ${\bf \hat{k}}=(k_x,k_y,k_z)$ is obtained using two uniform random
numbers $r_1$ and $r_2$ between 0 and 1 with the following prescription
\begin{eqnarray}
\mu &=& \cos\theta = 2r_1 - 1 \\
\nonumber
\phi &=& 2\pi r_2 \\
\nonumber
k_x &=& \sin\theta \cos \phi \\
\nonumber
k_y &=& \sin\theta \sin \phi \\
\nonumber
k_z &=& \cos\theta.
\end{eqnarray}\label{eq:wavevector}
Because scattered radiation is polarized in an anisotropic geometry, it is important to carry the polarization information. 
The four Stokes parameters $(I,Q,U,V)$ are necessary to describe the polarization state. An equivalent way is provided by considering the $2\times2$ density matrix defined by
\begin{equation}
\rho =
\begin{bmatrix}
{(I + Q)/2} & {(U + iV)/2} \\
{(U - iV)/2} & {(I - Q)/2} \\
\end{bmatrix}
.
\end{equation}.

In our simulation, the Stokes parameter $V$, representing circular polarization, is always set to zero because
no circular polarization develops from initially unpolarizd photons in electric dipole processes associated with Raman and Rayleigh scattering.
Initially unpolarized photons are described by a simple $\rho$ is given by
\begin{eqnarray}
\rho_{11} &=& 0.5 \\
\nonumber
\rho_{22} &=& 0.5 \\
\nonumber
\rho_{12} &=& \rho_{21} = 0\ .
\end{eqnarray}

The wavelength measured by an observer lying on the photon path is kept in the simulation. We assume that
the emitters are in random motion and also subject to the bulk motion associated with the grid. 
If we let the velocity of the emitter be ${\bf v}_{\rm emit}$
including bulk and random velocities,
the wavelength in the grid frame $\lambda_g$ is given by
\begin{equation}
\lambda_g = \lambda_i \left(  1 - {\frac{{\bf v}_{\rm emit} \cdot {\bf\hat k}}{c} } \right)
  \left(  1 + {\frac{{\bf v}_{\rm G} \cdot {\bf\hat k}}{c}} \right
),
\end{equation}
where ${\bf v}_{\rm G}$ is the velocity of the cell $(iX,iY,iZ)$.

\subsection{Journey of Photon: Optical Depth and Free Path }\label{sec:path}

\begin{figure*}
	\includegraphics[width=180mm]{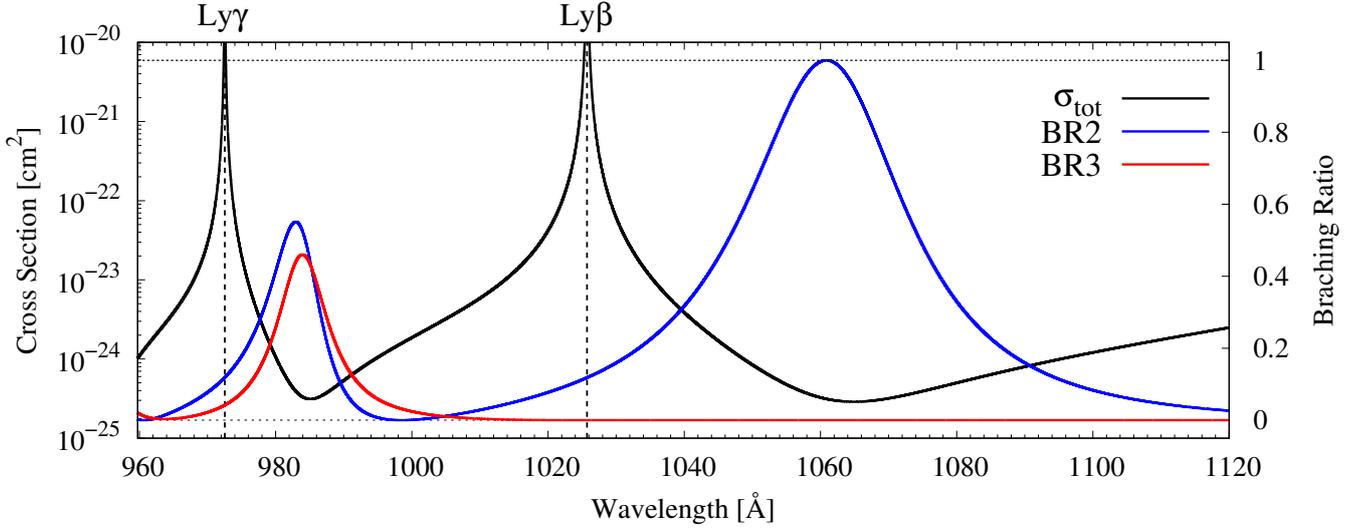}
	\caption{The total scattering cross section(black),
	the branching ratio to $n=2$ (blue) and $n=3$ (red) 
	computed in \cite{chang15}.
	}
	\label{fig:cross}
\end{figure*}

The generation of an initial photon is followed by the estimate of the free optical depth $\tau$ given by
\begin{equation}
\tau = - \ln r,
\end{equation}
where $r$ is a uniform random number between $0$ and $1$.
We adopt the method in \cite{seon09} to compute the free path and the next scattering position in grid-based 
geometry. In this section, we provide a brief description to compute the free path and determine the scattering position.

If the starting position ${\bf r}$ of the photon is in cell $A$, we measure the distance to the boundary
of the cell $d_A$ along the photon ray from the starting position. With $d_A$ we define the
scattering optical depth $\tau_A$ to the cell boundary by
\begin{equation}\label{eq:tau1}
\tau_A = \sigma_{\rm tot}(\lambda_A) n_{{\rm HI},A} d_A,
\end{equation}
where $\lambda_A$ is the wavelength in cell $A$.
Fig.~\ref{fig:cross} shows
the total scattering cross section $\sigma_{\rm tot}(\lambda)$ as a function of wavelength.
If $\tau_A>\tau$, then
the next scattering position ${\bf r}'$ is found in cell $A$ as follows
\begin{equation}
{\bf r'} = {\bf r} + \left( \frac{\tau}{\tau_A} \right) d_A {\bf \hat k}  .
\end{equation}

In the opposite case where $\tau_A < \tau$, the photon enters the neighboring cell $B$.
In this case, the same problem is obtained if we regard the entry point 
${\bf r}_e={\bf r}+d_A{\bf\hat k}$
as the new starting point of the photon with a new free optical path $\tau'$ reduced by $\tau_A$, or
\begin{equation}
{\tau}' = {\tau} - {\tau_A}. 
\end{equation} 
It should be noted that we are dealing with radiative transfer in a medium in motion. Therefore,
cell $B$ may move with a velocity different from that of cell $A$, in which case the photon wavelength along its
propagation direction may change on entering cell $B$ from cell $A$. Denoting by ${\bf v}_{\rm G,A}$ and ${\bf v}_{\rm G,B}$ 
the velocities of cells $A$ and $B$, respectively, we have
\begin{equation}
\lambda_{B} = \lambda_A \left( 1 - {\frac{{\bf v}_{\rm G,A} \cdot {\bf\hat k}}{c}} \right) 
\left( 1 + \frac{{\bf v}_{\rm G,B} \cdot {\bf\hat k}}{c} \right).
\end{equation}

\begin{figure}
	\includegraphics[width=90mm]{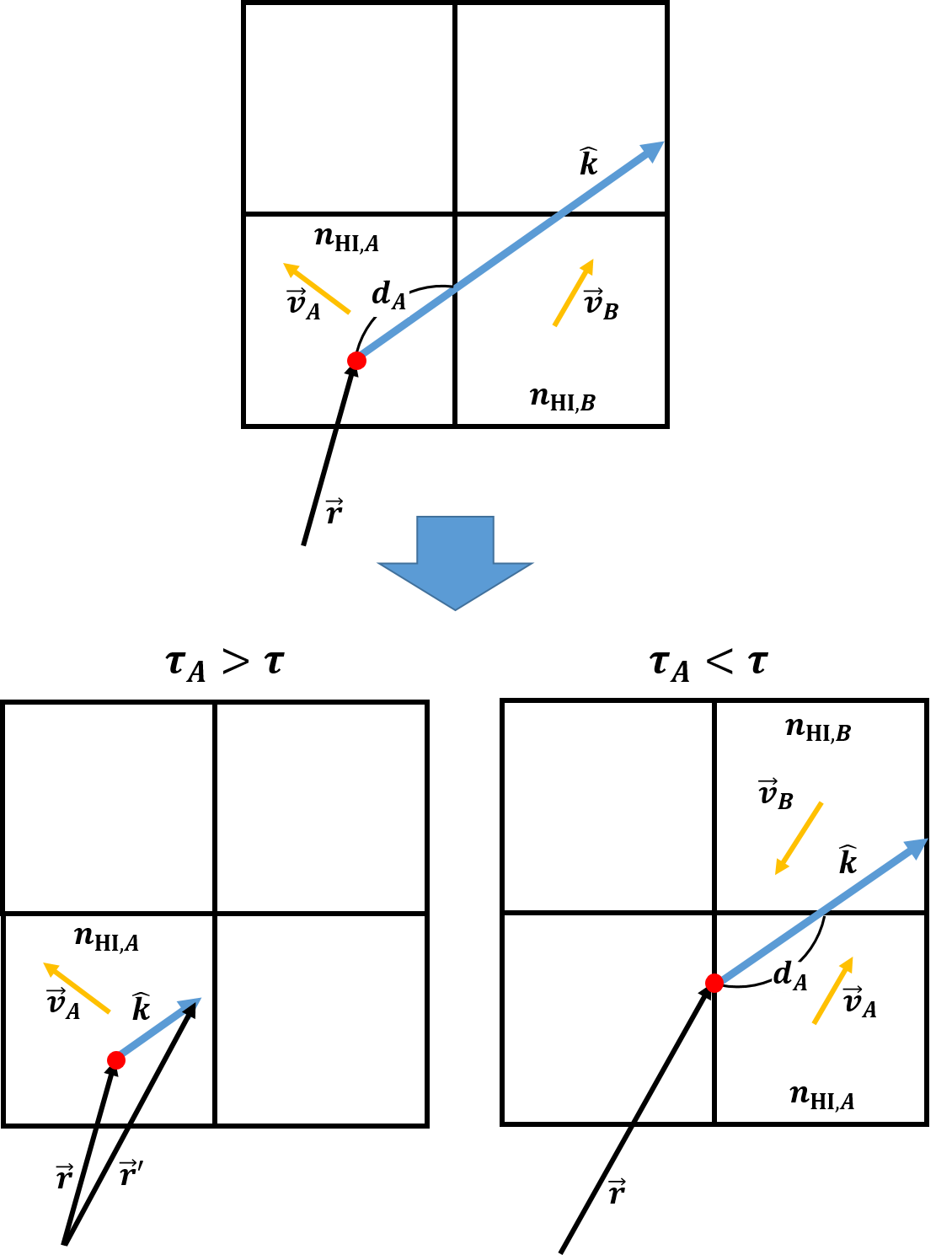}
	\caption{Schematic illustration of computing the next scattering position in the grid-based simulation.
	}
	\label{fig:grid}
\end{figure}

Iterations from Eq.~\ref{eq:tau1} are made with necessary updates ${\bf r}={\bf r}_e$, $\tau=\tau'$, and $\lambda_A = \lambda_B$ and
new naming of cell $B$ as cell $A$ until we have $\tau_A>\tau$.
Fig.~\ref{fig:grid} shows a schematic illustration of this procedure. 
In cases when a neighboring cell $B$ may not exist and $\bf r$ is outside the geometry,
the photon is regarded as Rayleigh-escaped.

\subsection{Rayleigh and Raman Scattering}\label{sec:scattering}

Raman scattering with atomic hydrogen with the initial and final electronic states being $1s$ and $2s$
shares the same the scattering phase function with Rayleigh scattering with atomic hydrogen in $1s$ 
\citep[e.g.][]{chang15}. It is also notable that the scattering is sufficiently far from resonance,
the scattering phase function is the same as that of Thomson scattering.
In this section, we describe the density matrix formalism with which
the polarization and the unit wavevector $\bf \hat k'$ of the scattered photon is determined  \citep{ahn15,chang17}.
We also describe the wavelength conversion and the line broadening associated with Raman scattering.

According to the density matrix formalism, the probability density of 
the scattered wavevector ${\bf\hat k'}=(\sin\theta'\cos\phi',\sin\theta'\sin\phi', \cos\theta')$ 
is given by
\begin{equation}
I'({\bf\hat k'}) = \rho'_{11} + \rho'_{22},
\end{equation}
where $\rho'_{ij}$ is defined by
\begin{equation}
\rho'_{ij}=\sum_{kl=1,2}(\hat{\epsilon}'_i\cdot \hat{\epsilon_k})\rho_{kl}(\hat{\epsilon_l}\cdot \hat{\epsilon_l}').
\end{equation}
Here, $\hat{\epsilon}_{1,2}$ and $\hat{\epsilon}_{1,2}'$ are 
the polarization basis vector associated with  $\bf \hat{k}$ and $\bf \hat{k}'$, respectively.
Specifically, $\hat\epsilon_1=(-\sin\phi,\cos\phi,0)$ and
$\hat\epsilon_2=(\cos\theta\cos\phi,\cos\theta\sin\phi, -\sin\theta)$ so that $\hat\epsilon_1$ represents
polarization in the direction perpendicular to the plane spanned by the photon wavevector and the
$z-$axis.

The components of the density matrix associated with the scattered radiation are related to those
of incident radiation by 
\begin{eqnarray}
\rho'_{11} &=& (\cos^2\Delta\phi)\rho_{11}
\nonumber\\
&-& (\cos\theta \sin2\Delta\phi)\rho_{12}
\nonumber\\
&+& (\sin^2\Delta\phi \cos^2\theta)\rho_{22}
\nonumber\\
\rho'_{12} &=& (\frac{1}{2} \cos\theta'\sin2\Delta\phi)\rho_{11}
\nonumber\\
&+& (\cos\theta \cos\theta' \cos2\Delta\phi + \sin\theta \sin\theta' \cos\Delta\phi)\rho_{12}
\nonumber\\
&-& \cos\theta(\sin\theta \sin\theta' \sin\Delta\phi + \frac{1}{2} \cos\theta \cos\theta' \sin2\Delta\phi)\rho_{22}
\nonumber\\
\rho'_{22} &=& (\cos^2\theta' \sin^2\Delta\phi)\rho_{11}
\nonumber\\
&+& \cos\theta'(2\sin\theta \sin\theta' \sin\Delta\phi + \cos\theta \cos\theta' \sin2\Delta\phi)\rho_{12}
\nonumber\\
&+& (\cos\theta \cos\theta' \cos\Delta\phi + \sin\theta \sin\theta')^2 \rho_{22},
\end{eqnarray}
where $\Delta \phi = \phi'-\phi$.
In the code, a selection is made for ${\bf\hat k}'$ from an isotropic distribution to compute $I'$. A new
random deviate $r'$ is compared with $I'$. If $r'<I'$, then the selection of ${\bf\hat k}'$ is accepted.
Otherwise, the process is iterated until the acceptance is obtained.
The scattered photon now becomes a new incident photon propagating to a new scattering site. This is
followed by necessary updates for the wavevector and the density matrix given by
\begin{eqnarray}
{\bf \hat k} &=& {\bf \hat k'} \\
\nonumber
\rho_{ij} &=& \rho'_{ij}.
\end{eqnarray}

The scattering type is determined by the branching ratio.
If an incident far UV photon is more energetic than Ly$\gamma$, the final states
available for Raman scattering include $2s$, $3s$ and $3d$ states.
For photons near Ly$\gamma$, the total cross section is given by
\begin{equation}
\sigma_{\rm tot} = \sigma_{1s} + \sigma_{2s} + \sigma_{3s+3d},
\end{equation}
where $\sigma_{1s}$, $\sigma_{2s}$, and $\sigma_{3s+3d}$ are 
the scattering cross sections corresponding to the final states $1s$, $2s$, and $3s+3d$, respectively.
Rayleigh branching ratio $BR1$ is given by
\begin{equation}
BR1 = \frac{\sigma_{1s}}{\sigma_{\rm tot}}.
\end{equation}
In a similar way,
Raman branching ratios corresponding to the final energy levels $n=2$ and $3$ are
\begin{equation}
BR2 = \frac{\sigma_{2s}}{\sigma_{\rm tot}}, \quad BR3 = \frac{\sigma_{3s+3d}}{\sigma_{\rm tot}},.
\end{equation}
In Fig.~\ref{fig:cross}, the blue and red solid lines represent $BR2$ and $BR3$, respectively.

The energy difference between the incident and Raman-scattered photons is the same as that
between the initial and final atomic states. This is translated into the relation between
the wavelengths $\lambda$ and $\lambda'$ of the Raman-scattered and incident photons, respectively, 
which is given by
\begin{equation}\label{eq:wavelength}
\frac{1}{\lambda} = \frac{1}{\lambda'} + \frac{1}{\lambda_{\rm res}},
\end{equation}
where $\lambda_{\rm res}$ is the wavelength corresponding to the energy difference between
the initial and final states.

One very important aspect in Raman scattering can be found in the conspicuous change in line width.
Differentiating Eq.~(\ref{eq:wavelength}), we have
\begin{equation}
\frac{d\lambda'}{\lambda'}= \left( \frac{\lambda'}{\lambda} \right)
\left( \frac{d\lambda}{\lambda} \right),
\end{equation}
from which we see immediately that the line width of Raman-scattered feature is
broadened by the factor $(\lambda'/\lambda)^2$.
For example, a typically observed line width of Raman-scattered O VI at 6825 \AA amounts to $\sim 30$~\AA\ 
whereas the far UV parent line O VI$\lambda\lambda$1032 exhibits a line width $\sim 1$~\AA\ in many symbiotic
stars.

Due to the line broadening effect, far UV radiation near Lyman series of hydrogen will be considerably diluted
and redistributed around Balmer emission lines to appear as broad wings \citep[e.g.][]{yoo02,chang15,chang18}. 
Another important consequence of the line broadening effect is that the line profiles of Raman-scattered
features mainly reflect the relative motion between the far UV emitters and the neutral scatterers and quite 
independent of the observer's line of sight \citep{heo16,choi20}.

\section{Code Test}

\begin{figure*}
	\includegraphics[width=170mm]{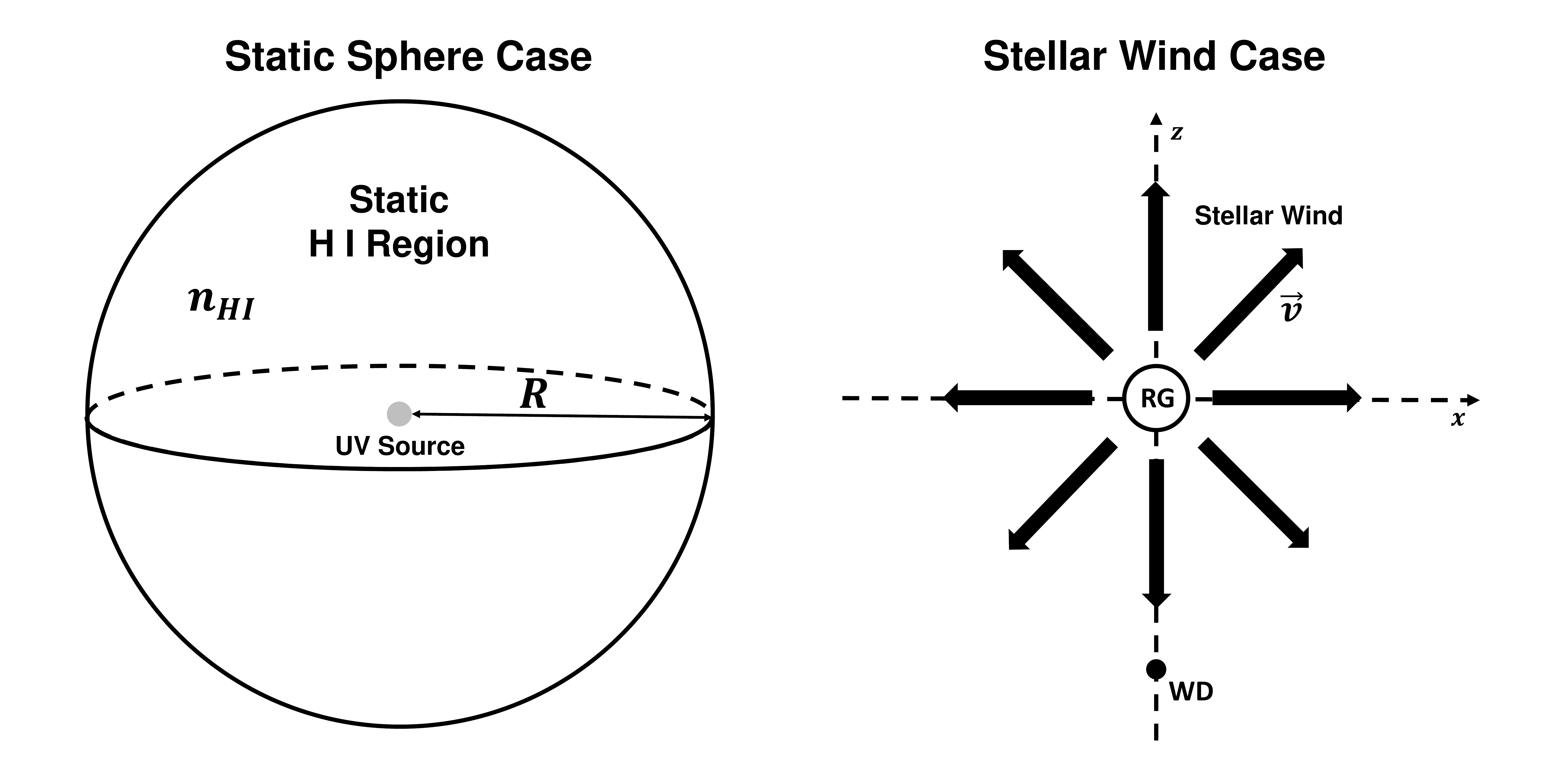}
	\caption{Schematic illustrations of two cases to test the code.
	}
	\label{fig:shell_model}
\end{figure*}

As a check of our code, we present our simulation results for two exemplary cases. 
The first example is a static spherical H I region surrounding a far UV
continuum source located at the center, in which Balmer wings are formed through Raman scattering. 
Analytic solutions are available for this case,
against which our result obtained from STaRS is compared. 
The second case is reproduction of the result of \cite{lee97b}, who investigated Raman scattering
of O VI in symbiotic stars. 
In this case, the H I region is an expanding spherical wind around the giant component.
Schematic illustrations of the two cases are shown in  Fig.~\ref{fig:shell_model}.

\subsection{Formation of Balmer Wings in a Static Spherical H~I Region}

The central point-like far UV source surrounded by a spherical H I region with radius $R$ is characterized by a flat continuum. 
The static neutral region is assumed to be of
uniform H I density $n_{\rm HI}$. We fix the radial column density $N_{\rm HI}$ defined by
\begin{equation}
N_{\rm HI}=n_{\rm HI} R = 10^{23}{\rm\ cm^{-2}},
\end{equation}
and vary the number of cells. We set $N_{xyz}=N_x=N_y=N_z$, 
so that the total number of cells is given by $N_{xyz}^3$.
We generate $10^7$ photons for each simulation.
The initial photons are generated at the center of the H I sphere in accordance with
	the emisivity $j_{e}$ given by the three dimensional Dirac delta function 
	\begin{equation}
j_{e}(\lambda,x,y,z) =
{\lambda \over {\lambda_{max}-\lambda_{min}}} \delta^{(3)}({\bf r}), 
\end{equation}
where $\lambda_{max}$ and $\lambda_{min}$ are the maximum and minmum wavelengths of initial photons.

\begin{figure*}
	\includegraphics[width=170mm]{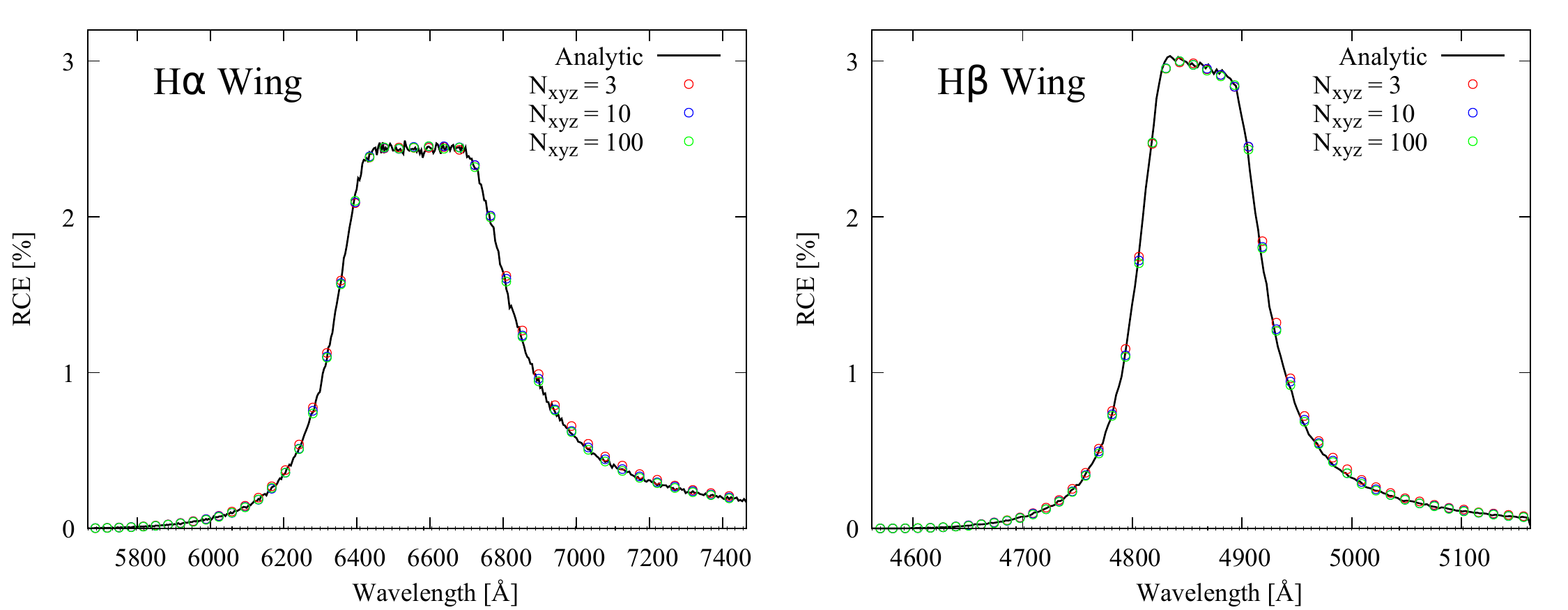}
	\caption{Raman conversion efficiency of H$\alpha$ (left panel) and H$\beta$ (right panel) broad wings of the static spherical case.
	The black solid lines are the spectra by the analytic method.
	The $x$ axis is the wavelength.
	The red, blue, and green open circles are the spectra by STaRS for $N_{xyz}= 3$, 10, and 100.
	}
	\label{fig:shell_spec}
\end{figure*}

Fig.~\ref{fig:shell_spec} shows optical spectra formed 
through Raman scattering for the cases of $N_{xyz}=3, 10$ and 100. 
The vertical axis shows the Raman conversion efficiency ($RCE$)
defined as the number ratio per unit wavelength of incident far UV photons and Raman-scattered optical photons. 
The left and right panels show Balmer wings formed through Raman scattering
around H$\alpha$ and H$\beta$, respectively.
The solid lines represent the analytic solutions and open circles 
show simulation results obtained using STaRS.
The analytic solutions are obtained from the non-grid based simulation in \cite{chang15,chang18}.

Far from the line centers, the simulation results for $N_{xyz}=3$ are slightly higher
than the analytic solutions. 
Other than this, the agreement is fairly good, 
indicating little dependence on $N_{xyz}$ in the case of a static H I region.
The volume of the H I region with $N_{xyz}=3$ is larger than the sphere with the radius $R$
as the H~I density associated with a cell is determined by the central position $(X_C,Y_C,Z_C)$ of the cell.
When $(X_C^2 +Y_C^2 +Z_C^2)^{1/2}$ is smaller than $R$, the H~I density of the cell is assigned to be $n_{HI}$.
Otherwise, the density is assigned to be zero.
The number ratio between Raman scattered photons near H$\alpha$ and the total initial photons near Ly$\beta$ is
20.77 \% for the analytic solution, whereas the simulations give 21.17, 20.91, and  20.77 \% for $N_{xyz} = 3$, 10, and 100,
respectively.

In Fig.~\ref{fig:image_ha}, we present the polarimetric data of Raman scattered H$\alpha$ projected to the celestial sphere.
The surface brightness, the degree of polarization $p$, and the direction of polarization are shown in the left, middle and the right panels, respectively.
Here, in terms of the Stokes parameters,
the degree of polarization $p$ and the position angle $\phi_p$ are given by
\begin{equation}
    p = \frac{\sqrt{Q^2 +U^2}}{I}, \quad \phi_p = \frac{1}{2} \tan^{-1}\left( \frac{U}{Q} \right).
\end{equation}
In the left panels, the surface brightness is shown in logarithmic scale.
With our choice of a rather large value of $N_{\rm HI}$,
a considerable fraction of Raman-scattered photons are formed near the source, which
leads to excellent agreement between the analytic results and those obtained using STaRS.
One may notice increase in smoothness of the surface brightness as $N_{xyz}$ increases.
It is also noticeable that $p$ becomes large with increasing distance from the center.
The concentric polarization pattern reflects the spherically symmetric scattering geometry.
In this particular case, $N_{xyz}>10$ appears to be sufficient to describe the the analytic result of the static medium.

\subsection{Raman Scattering of O VI in Expanding H I Region}

\cite{lee97b} presented their basic study of line formation of Raman O VI  in a symbiotic star consisting of a white dwarf and a mass losing giant.
The O VI emission region
near the white dwarf component was assumed to be a point source broadened thermally with $T=10^4\, \rm K$.
For simplicity, the slow stellar wind from the red giant component is assumed to be entirely neutral ignoring
the photoionization by the white dwarf. 
The orbital separation is $10R_*$ of which $R_*$ is the radius of the red giant.
The positions of the white dwarf and the red giant are $(0,0,10R_*)$ and $(0,0,0)$, respectively.
The initial photons are generated at the position of the white dwarf according to
	the emissivity $j_{e}$ given by
	\begin{eqnarray}
		&&j_{e}(\lambda,x,y,z) = \\
		\nonumber
		&&{1 \over {\sigma_{th} \sqrt{2\pi}}}  \exp\left[{(\lambda - \lambda_{1032})^2 \over {2 \sigma_{th}^2}} \right] \delta^{(3)}(x,y,z-10R_*) 
	\end{eqnarray}
	where $\lambda_{1032}$ is the center wavelength of O VI $\lambda$ 1032 and
	$\sigma_{th} \sim 0.008{\rm\ \AA}$ is the thermal width of O VI $\lambda$ 1032 with $T = 10^4 K$.

The velocity $\bf v(r)$ and H I number density $n({\bf r})$ are the functions of the distance from the red giant $r=|\bf r|$
 are given by
\begin{eqnarray}
    {\bf v}({\bf r}) &=& v_{\infty}(1 - R_*/r) \frac{\bf r}{ r}\\
    \nonumber
    \quad
    n({\bf r}) &=& n_0 (R_*/r)^{2} (1 - R_*/r)^{-1},
\end{eqnarray}
where $n_0$ is the characteristic number density defined in Eq.~2.11 of \cite{lee97b}
and $v_{\infty}$ is the terminal velocity.
As a second code test case, we revisit Raman O VI formation illustrated in Figs.~1 and 6 of \cite{lee97b}.
Fixing $v_\infty=20\, \rm km\, s^{-1}$, we consider the three values of
$\tau_0 = 0.5 $, 1, and 10, where $\tau_0 = n_0 R_* \sigma_{\rm tot}$.
We generate $10^9$ photons for each simulation.

Fig.~\ref{fig:lee_1997} shows the spectra and the Stokes parameter
$P=Q/I$ for Raman scattered O VI at 6825 \AA.
It should be noted that the integrated $U$ vanishes due to the axial symmetry about the $z$ axis.
Therefore, the signed ratio $Q/I$ represents the degree and direction of polarization, where
a positive  and a negative $Q$ correspond to  the polarization in direction perpendicular and parallel to the $z$ axis, respectively.
The observer's line of sight lies in $x$-$y$ plane.
Noting that it is perpendicular to the symmetry axis,
maxium polarization can be developed along this direction or the symmetry $z-$axis.
We collecte the photons escaping toward the observer.
The fraction of the collected Raman photons near 6825 \AA\, is
0.11, 0.21, and 1.82 \% for $\tau_0 = 0.5$, 1, and 10, respectively.

The line profiles obtained using STaRS differ slightly from those \cite{lee97b} presented.
In particular, 
the red peaks obtained from STaRS are more enhanced than those presented by \cite{lee97b},
which is attributed to the geometrical truncation adopted by them.
The difference is in the range of 10-15 per cent with respect to the red peak.
We find that the agreement gets better where the full range of the scattering region is taken into account.
In the bottom panels, we find overall agreement in the polarization behaviors.
The noisy features with $\Delta V > 200 \rm\, km\, s^{-1}$ are attributed to the small number statistics of 
collected photons.
Blue photons are scattered mostly in the compact and dense region between the red giant and white dwarf, resulting in
development of strong polarization in the direction perpendicular to the $z$ axis.
In contrast, red photons are scattered mainly in a quite extended region near the red giant, 
leading to weak polarization and enhanced line flux.

\begin{figure*}
	\includegraphics[width=165mm]{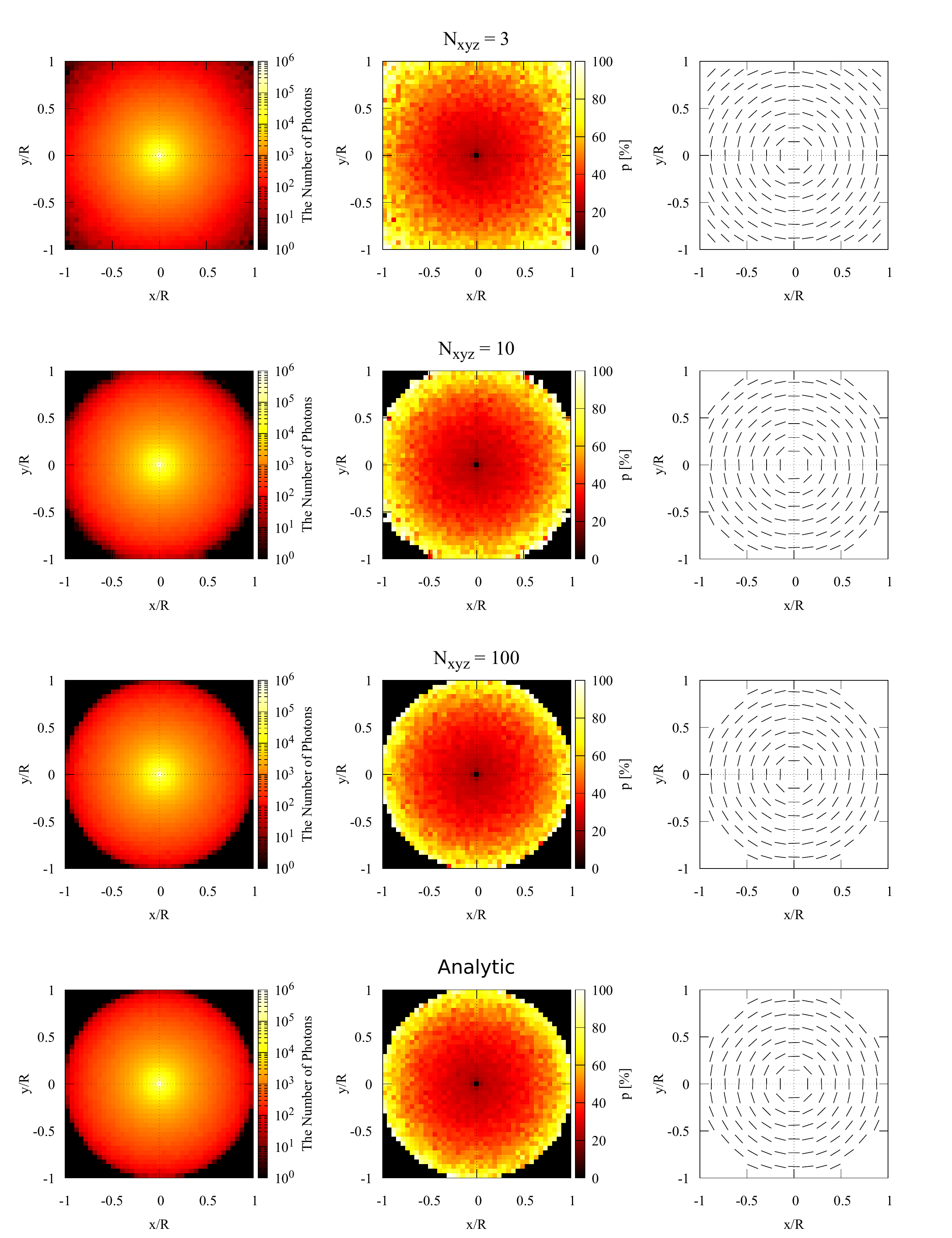}
	\caption{The surface brightness (left), the degree of polarization (center), and the direction of polarization (right) of the projected H$\alpha$ photons.
    The panels in the first, second, and third rows represent the results obtained using STaRS for $N_{xyz}= 3$, 10, and 100. The bottom panels represent the result obtained using an analytic method.
	}
	\label{fig:image_ha}
\end{figure*}

\begin{figure*}
	\includegraphics[width=170mm]{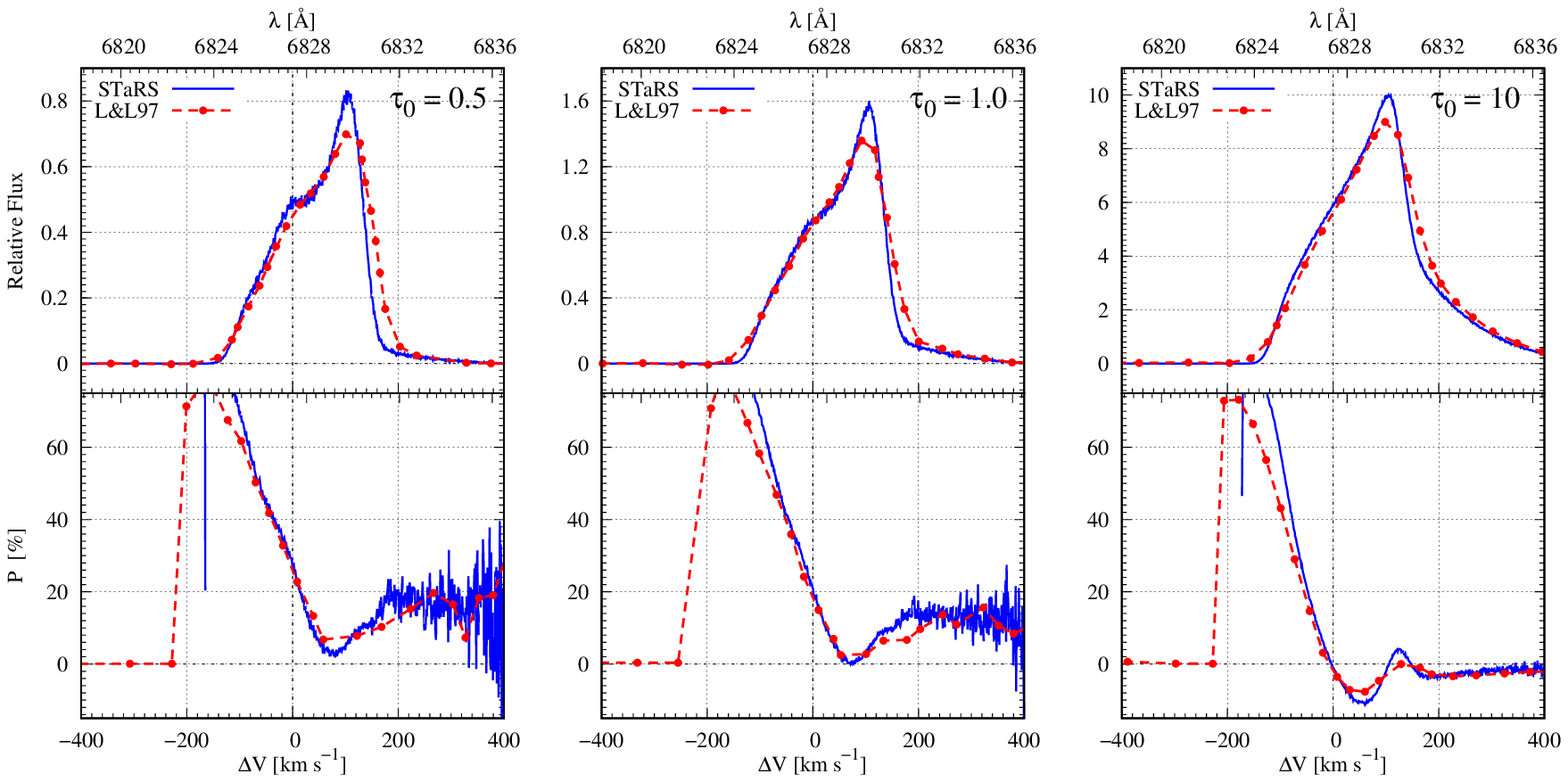}
	\caption{Line formation of Raman O VI 6825 features in an expanding H I region. 
	The blue solid lines are computed by 'STaRS'. 
	The red dashed lines are the results of Fig.~6 in \cite{lee97b}.
	}
	\label{fig:lee_1997}
\end{figure*}

Fair agreement shown in  Fig.~\ref{fig:shell_spec} and \ref{fig:lee_1997} demonstrates
that STaRS has been well-tested. Furthermore, as illustrated in
Fig.~\ref{fig:image_ha}, STaRS is capable of study of radiative transfer for spectropolarimetric imaging observations.

\section{Summary}

We have developed a 3D grid-based Monte Carlo code 'STaRS' for radiative transfer through Raman
and Rayleigh scattering, which can be mainly used to investigate line formation of Raman-scattered
features in a thick neutral region illuminated by a strong far UV emission source. Favorable
conditions for Raman scattering with atomic hydrogen are easily met in symbiotic stars, young planetary nebulae and active galactic nuclei. 
Through a couple of tests, we have successfully demonstrated that 'STaRS' is a flexible
code to deal with radiative transfer in a thick neutral media yielding multidimensional
spectropolarimetric and imaging data.
'STaRS' is easily accessed in Github 'https://github.com/csj607/STaRS'.

\acknowledgments

This research was supported by the Korea Astronomy and Space Science Institute 
under the R\&D program (Project No. 2018-1-860-00) super-vised 
by the Ministry of Science, ICT and Future Planning. 
This work was also supported by a National Research Foundation of Korea (NRF) grant
funded by the Korea government (MSIT; No. NRF-2018R1D1A1B07043944).
Seok-Jun is very grateful to Dr. Kwang-Il Seon for his help
in adoption of grid-based technique.



\end{document}